\documentclass[twocolumn, times, tighten,twocolappendix]{aastex631}
\usepackage{graphicx,multirow,hyperref,url,color,xspace}
\usepackage{amsmath,amssymb}

\newcommand{\msun}{{\rm M}_{\sun}}

\newcommand{\hxmt}{{\textit{Insight-HXMT}}\xspace}

\newcommand{\nustar}{{\textit{NuSTAR}}\xspace}
\newcommand{\nicer}{NICER\xspace}
\newcommand{\integral}{{\textit{INTEGRAL}}\xspace}
\newcommand{\source}{{MAXI J1820+070}\xspace}
\newcommand{\g}{{$\gamma$}}

\newcommand{\appropto}{\mathrel{\vcenter{
  \offinterlineskip\halign{\hfil$##$\cr
    \propto\cr\noalign{\kern2pt}\sim\cr\noalign{\kern-2pt}}}}}

\defcitealias{Zdziarski21c}{Z21c}
\newcommand{\zdz}{{\citetalias{Zdziarski21c}}\xspace}

\defcitealias{You21}{Y21}
\newcommand{\bei}{{\citetalias{You21}}\xspace}

\usepackage{ulem}

\begin{document}

\title{\hxmt, \nustar and \integral Data Show Disk Truncation in the Hard State of the Black-Hole X-Ray Binary MAXI~J1820+070}
\shorttitle{MAXI J1820+070 with \hxmt, \nustar and \integral}

\author{Andrzej A. Zdziarski}
\affiliation{Nicolaus Copernicus Astronomical Center, Polish Academy of Sciences, Bartycka 18, PL-00-716 Warszawa, Poland; \href{mailto:aaz@camk.edu.pl}{aaz@camk.edu.pl}}
\author{Bei You}
\affiliation{School of Physics and Technology, Wuhan University, Wuhan 430072, China}
\affiliation{Astronomical Center, Wuhan University, Wuhan 430072, China}
\author{Micha{\l} Szanecki}
\affiliation{Faculty of Physics and Applied Informatics, {\L}{\'o}d{\'z} University, Pomorska 149/153, PL-90-236 {\L}{\'o}d{\'z}, Poland}
\author{Xiao-Bo Li}
\affiliation{Key Laboratory of Particle Astrophysics, Institute of High Energy Physics, Chinese Academy of Sciences, Beijing 100049, China}
\author{Mingyu Ge}
\affiliation{Key Laboratory of Particle Astrophysics, Institute of High Energy Physics, Chinese Academy of Sciences, Beijing 100049, China}

\shortauthors{Zdziarski et al.}

\begin{abstract}
We study X-ray and soft gamma-ray spectra from the hard state of the accreting black-hole binary MAXI J1820+070. We perform analysis of joint spectra from \textit{HXMT}, \textit{NuSTAR} and \textit{INTEGRAL}. We find an overall agreement between the spectra from all three satellites. Satisfactory fits to the data require substantial spectral complexity, with our models including two Comptonization regions and their associated disk reflection, a disk blackbody and a narrow Fe K$\alpha$ line. Our fits confirm the presence of the truncation of the reflecting optically-thick disk at least at $>$10 gravitational radii. However, we find that the \textit{HXMT} data alone cannot significantly constrain the disk inner radii. 
\end{abstract}

\section{Introduction}
\label{intro}

The knowledge of the geometry of accretion flows onto black holes (BHs) is crucial for understanding the physics of accretion. While there is largely a consensus that the standard disk model of geometrically-thin, optically-thick accretion disk \citep{SS73,NT73} extending down to the innermost stable circular orbit (ISCO) applies to the soft spectral state of accreting BH binaries, there is an ongoing controversy regarding the geometry of accretion in the hard spectral state. The paradigm dominant for a number of years that the inner disk is truncated and replaced by a hot flow (e.g., \citealt{DGK07}) has been questioned in papers claiming the disk extends down almost to the ISCO also in the hard state, see \citet{Bambi21} for a recent review. 

Here, we address this question for \source, a bright transient accreting BH binary, whose outburst was discovered in 2018 \citep{Tucker18,Kawamuro18}. The presence of a disk extending almost to the ISCO in its hard state was advocated in \citet{Kara19}, \citet{Buisson19} and \citet{Wang21}. On the other hand, studies by \citet{Wang20_HXMT}, \citet{Axelsson21}, \citet{Dzielak21}, \citet{DeMarco21}, \citet{Marino21}, \citet{Zdziarski21b}, \citet{Zdziarski21c}, hereafter \zdz, and \citet{Kawamura21} found the disk to be significantly truncated over various stages of the hard state.

\citet{You21}, hereafter \bei, analyzed broad energy spectra (2--200\,keV) in the hard state from {\it Hard X-ray Modulation Telescope} (\hxmt; \citealt{Zhang14}) for \source in the 2018 outburst. It was found that the reflection fraction, defined as the ratio of the coronal flux that illuminates the disk to that emitted outside, showed an overall decrease with the decreasing hardness of the X-ray spectrum. The latter decrease, in turn, was found \citep{Kara19,DeMarco21} to be strongly correlated with the decreasing time lag of soft X-rays with respect to hard ones observed by the Neutron star Interior Composition ExploreR (\nicer; \citealt{Gendreau16}). \bei interpreted this effect as indicating that the corona is relativistically outflowing with the outflow velocity increasing with the decreasing coronal scale height. However, the spectral fits of \hxmt alone cannot put constraint on the inner radius of the disc. Here, we follow up on the studies of \bei and \zdz, and combine the data from \hxmt used in \bei with those from {\it Nuclear Spectroscopic Telescope Array} (\nustar; \citealt{Harrison13}), the Spectrometer on \integral (\citealt{Roques03}; SPI), and the \integral Soft Gamma-Ray Imager (\citealt{Lebrun03}; ISGRI) used in \zdz. We apply two different X-ray spectroscopy models to the joint data, with the main goal of estimating the truncation radius of the disk. Also, we estimate other parameters of the X-ray source. 
 
The parameters of binary relevant to our study are as follows. The distance is $D\approx 2.96\pm 0.33$\,kpc based on a radio parallax \citep{Atri20}, and $D\leq 3.11\pm 0.06$\,kpc based on the proper motion of the moving ejecta during the hard-to-soft state transition \citep{Wood21}. The inclinations of the binary and the radio jet are $i_{\rm b}\approx 66\degr$--$81\degr$ \citep{Torres20}, $i_{\rm j}\approx 64\degr\pm 5\degr$ \citep{Wood21}, respectively, and the BH mass is $M\approx (5.95\pm 0.22)\msun/\sin^3 i_{\rm b}$ \citep{Torres20}.

\section{Observations and data reduction}
\label{data}

\setlength{\tabcolsep}{4pt}
\begin{table*}\centering
\caption{Contemporaneous observations of \source with \hxmt, \nustar and \integral in the hard state  
}
\begin{tabular}{lccccccccccc}
\hline
Epoch & \hxmt & Start time &Exp.\ LE, ME& SPI Start time & ISGRI Start time & Exp.\ SPI &\nustar & Start time & Exp.\ A \\
& Obs.\ ID & End time&Exp.\ HE& End time & End time &Exp.\ ISGRI &Obs.\ ID& End time&  Exp.\ B \\
\hline
1 &P011466100402--3&58201.514 & 3688, 4049&58201.555& 58201.544 &13604 &90401309008 &58201.526&3046  \\
&&58201.747& 5599&58201.757&58201.757&8796  &&58201.766 &3214\\
2 &P011466101202--4 & 58212.203&3800, 3859 &58212.181 &58212.181 & 28507 &90401309012 &58212.200&12334   \\
&&58212.541&5402 &58212.606&58212.393&8305&&58213.177&12964\\
\hline
\end{tabular}
\tablecomments{
The times are given in MJD, and the exposures are effective in seconds. }
\label{log}
\end{table*}

For our analysis, we have chosen epochs of the hard state of \source for which simultaneous data from all of \hxmt, \nustar and \integral are available. We have found only two such epochs. They are the same as epochs 1 and 2 in \zdz, which work analyses the \nustar and \integral data only. In the current study, we add to them the corresponding \hxmt spectral data. The two epochs correspond to the beginning of a plateau phase on the count-rate/hardness plot, see Fig.\ 1 of \citet{Zdziarski21b}. As shown in that work, see their table 2, the bolometric luminosity during that part of the plateau phase was almost constant, at $\approx$15\% of the Eddington luminosity (assuming $D=3$\,kpc and $M=8\msun$).

The \integral data (see Table \ref{log}) are from SPI and ISGRI, which jointly cover the 23--650\,keV range, and are the same as those in \zdz. The analysis of the SPI data is described in \citet{Roques19}. Data reduction and spectral extraction for ISGRI was done using the {\sc osa} v.\ 11.1 software \citep{Courvoisier03}. The SPI and ISGRI data include 0.5\% and 1\% systematic errors, respectively. 

The data from \nustar (3--79\,keV) were reduced with {\sc heasoft} v.6.25, the {\tt NuSTARDAS} pipeline v.1.8.0, and {\tt CALDB} v.20200912 from the source region of a $60''$ circle centered on the peak brightness. The data were grouped to the signal-to-noise ratio $\geq$50 below 69\,keV and no grouping has been applied at higher energies. Following previous spectral studies of the \nustar data, e.g., \citet{Buisson19}, we have applied no systematic errors to the data. The data consist of those from two focal plane modules, A and B, see Table \ref{log}. 

The \hxmt data are those extracted by \bei. The chosen data sets are listed in Table \ref{log}. The low, medium and high energy detectors, LE, ME, and HE, are used in the 2--10, 10--30, and 30--200\,keV ranges, respectively. The individual \hxmt data overlapping with those of \nustar have been merged, as described in Table \ref{log}. The data were grouped according to the recommendation of the \hxmt team. In order to directly study the spectral calibration of the \hxmt detectors, we have applied no systematic errors to these data. See \citet{Li20} for estimates of the actual systematic errors.

\setlength{\tabcolsep}{5pt}
\begin{table}
\caption{The results of spectral fitting the \nustar+\integral data alone in the 3--650\,keV range
}
   \centering\begin{tabular}{ccccc}
\hline
Component & Parameter & Epoch 1 & Epoch 2 \\
\hline
ISM absorption & $N_{\rm H}$ $[10^{21}]$\,cm$^{-2}$ & \multicolumn{2}{c}{1.4f}\\
\hline
Joint constraints & $i$ $[\degr$] & $59^{+3}_{-0}$ & $67^{+4}_{-1}$\\
& $Z_{\rm Fe}\, [\sun]$ & $3.6^{+0.5}_{-1.7}$ & $1.7^{+0.2}_{-0.1}$\\
\hline
Disk & $kT_{\rm in}$ & $0.5^{+0.1}_{-0.1}$ & $0.4^{+0.1}_{-0.1}$\\
and a narrow & $N_{\tt diskbb}\,[10^{3}]$ & $7^{+19}_{-6}$ & $10^{+5}_{-5}$\\
6.40\,keV line & $N_{{\rm Fe\,K}\alpha}\,[10^{-3}]$ & $4^{+1}_{-1}$ & $2^{+1}_{-1}$\\
\hline
Thermal  & $y_{\rm th}$ & $0.62^{+0.01}_{-0.01}$ & $0.58^{+0.02}_{-0.01}$\\
Comptonization &$\Gamma_{\rm th}$ & $1.66^{+0.01}_{-0.02}$ & $1.72^{+0.04}_{-0.01}$ \\
and reflection & $kT_{\rm e,th}$ [keV] & $11^{+1}_{-1}$ & $11^{+1}_{-1}$\\
&$R_{\rm in}\, [R_{\rm g}]$ & $13^{+5}_{-4}$ & $17^{+3}_{-5}$\\
& ${\cal R}_{\rm th}$ & $1.5^{+0.2}_{-0.3}$ & $0.9^{+0.2}_{-0.4}$\\
& $\log_{10} \xi_{\rm th}$ & $4.3^{+0.1}_{-0.1}$ & $3.9^{+0.1}_{-0.1}$\\
& $N_{\rm th}$ & $1.51^{+1.86}_{-0.03}$ & $2.26^{+0.01}_{-0.04}$\\
\hline 
Hybrid &$y_{\rm h}$ & $1.18^{+0.01}_{-0.01}$ & $0.93^{+0.05}_{-0.04}$\\
Comptonization &$\Gamma_{\rm h}$  & $1.26^{+0.01}_{-0.01}$ & $1.27^{+0.04}_{-0.03}$\\
and reflection& $kT_{\rm e,h}$ [keV] & $23^{+1}_{-1}$ & $19^{+1}_{-1}$\\
&$\gamma_{\rm min}$ & $1.30^{+0.02}_{-0.05}$ & $1.09^{+0.29}_{-0.01}$\\
&$p$ & $2.70^{+0.37}_{-0.09}$ & $3.45^{+0.14}_{-0.15}$\\
&$\Delta R\, [R_{\rm g}]$ & $20^{+11}_{-6}$ & $35^{+13}_{-9}$\\ 
& ${\cal R}_{\rm h}$ & $0.74^{+0.25}_{-0.08}$ & $0.72^{+0.06}_{-0.03}$\\
& $\log_{10} \xi_{\rm h}$ & $2.3^{+0.2}_{-0.3}$ & $1.7^{+0.2}_{-1.7}$\\
& $N_{\rm h}$ & $0.47^{+0.30}_{-0.04}$ & $0.55^{+0.05}_{-0.11}$\\
\hline
& $\chi_\nu^2$  & 920/789 & 1634/1304\\
\hline
\end{tabular}
\tablecomments{The {\tt reflkerr} model, Equation \ref{reflkerr}, is used. $Z_{\rm Fe}$ is the Fe abundance in solar units, the inner radius of the outer, weakly ionized, reflection is $R_{\rm tr}=R_{\rm in}+\Delta R$, $R_{\rm out}=10^3 R_{\rm g}$, $N_{\rm th,h}$ is the flux density of a Compton component @1\,keV, $y_{\rm th,h}$ is the Compton parameter (calculated accurately for spherical geometry, see \zdz), $\Gamma_{\rm th,h}$ is the power-law index fitted to a Compton component in the 2--10\,keV range (not a free parameter), and ${\cal R}_{\rm th,h}$ is the reflection fraction. }
\label{t_reflkerr}
\end{table}

\setlength{\tabcolsep}{5pt}
\begin{table}
\caption{The results of spectral fitting the \nustar+\integral data alone in the 3--150\,keV energy range
}
   \centering\begin{tabular}{ccccc}
\hline
Component & Parameter & Epoch 1 & Epoch 2 \\
\hline
ISM absorption & $N_{\rm H}$ $[10^{21}]$\,cm$^{-2}$ & \multicolumn{2}{c}{1.4f}\\
\hline
Joint constraints & $i$ $[\degr$] & $59^{+3}_{-0}$ & $59^{+7}_{-0}$\\
& $Z_{\rm Fe}\, [\sun]$ & $1.5^{+0.9}_{-0.1}$ & $2.0^{+0.7}_{-0.3}$\\
\hline
Disk & $kT_{\rm in}$ & $0.4^{+0.2}_{-0.1}$ & $0.4^{+0.1}_{-0.1}$\\
and a narrow & $N_{\tt diskbb}\,[10^{3}]$ & $49^{+1400}_{-39}$ & $13^{+69}_{-9}$ \\
6.40\,keV line & $N_{{\rm Fe\,K}\alpha}\,[10^{-3}]$ & $4^{+1}_{-1}$ & $2^{+1}_{-1}$ \\
\hline
Soft  &$\Gamma_{\rm s}$ & $1.69^{+0.01}_{-0.03}$ & $1.71^{+0.01}_{-0.01}$ \\
Comptonization & $kT_{\rm e,s}$ [keV] & $28^{+34}_{-6}$ & $11^{+1}_{-1}$\\
and reflection &$R_{\rm in,s}\, [R_{\rm g}]$ & $12^{+4}_{-4}$ & $63^{+58}_{-26}$\\
& ${\cal R}_{\rm s}$ & $0.7^{+0.2}_{-0.2}$ & $0.7^{+0.5}_{-0.2}$\\
& $\log_{10} \xi_{\rm s}$ & $3.8^{+0.4}_{-0.1}$ & $4.0^{+0.1}_{-0.1}$\\
& $N_{\rm s}\;[\times 10^{-2}]$ & $7.7^{+0.1}_{-0.2}$ & $5.7^{+0.1}_{-0.1}$\\
\hline 
Hard  &$\Gamma_{\rm h}$  & $1.43^{+0.04}_{-0.02}$ & $1.20^{+0.04}_{-0}$\\
Comptonization& $kT_{\rm e,h}$ [keV] & $43^{+19}_{-17}$ & $26^{+1}_{-1}$\\
and reflection
&$R_{\rm in,h}\, [R_{\rm g}]$ & $39^{+13}_{-8}$ & $47^{+13}_{-10}$\\ 
& ${\cal R}_{\rm h}$ & $1.26^{+0.39}_{-0.04}$ & $0.9^{+0.1}_{-0.2}$\\
& $\log_{10} \xi_{\rm h}$ & $0^{+2.4}$ & $1.8^{+0.4}_{-1.8}$\\
& $N_{\rm h}\;[\times 10^{-2}]$ & $5^{+3}_{-2}$ & $0.2^{+0.1}_{-0.1}$\\
\hline
& $\chi_\nu^2$  & 905/772 & 1585/1289\\
\hline
\end{tabular}
\tablecomments{The {\tt relxillCp} model, Equation \ref{relxill}, is used. }
\label{t_relxill}
\end{table}
 
\section{Spectral Fits}
\label{fits}

Here, we further develop spectral models of \citet{Zdziarski21b} and \zdz fitted to X-ray and soft \g-ray data from \source in the hard state. In \citet{Zdziarski21b}, we fitted four observations by \nustar, which cover the 3--79\,keV energy range. We have found that those spectral data require the presence of two different, softer and harder, Comptonization components, and two corresponding reflection components, one strongly relativistically blurred and strongly ionized, and one weakly blurred and weakly ionized. No disk blackbody component was included. The Comptonization was modelled as by purely thermal electrons. The harder component dominated the broad-band X-ray flux in all cases. Our proposed geometry (see fig.\ 4 in \citealt{Zdziarski21b}) was that of a hot flow within the disk inner truncation radius, $R_{\rm in}$, with a large scale height responsible for the harder Comptonization, and a coronal flow covering the disk between $R_{\rm in}$ and a transition radius, $R_{\rm tr}$. The softer primary component reflected from the disk underneath the corona, while the harder one reflected from the radii $>\! R_{\rm tr}$. The truncation radii were found to be in the range of $R_{\rm in}\approx 10$--$10^2 R_{\rm g}$, where $R_{\rm g}$ is the gravitational radius. As pointed out in \citet{Zdziarski21b}, the harder component may instead be outflowing and forming a slow sheath of the jet. 

In \zdz, we studied two sets of simultaneous observations of \source in the hard state by \nustar and \integral. The joint data covered the 3--650\,keV range. Our major finding was that Comptonization by purely thermal electrons was not sufficient to account for the emission at $\gtrsim$100\,keV, while the high-energy spectrum formed a smooth continuation of that at lower energies. The full spectra were then well fitted by the same model as above except that the hard Comptonization was by hybrid electrons, i.e., Maxwellian with a high energy tail. The electron temperatures were found rather low, $kT_{\rm e}\approx 10$--30\,keV. The truncation radius was found to be relatively high, $R_{\rm in}\approx 20$--$30 R_{\rm g}$. Also, the average spectrum from all \nustar and \integral observations in the hard state was studied. It was similar in shape to those of the two individual observations, but with a much better statistics at high energies, which allowed a reliable calculation of the rate of e$^\pm$ pairs. That rate was found to be quite high, but balanced by pair annihilation, leading to the equilibrium relative pair density being low, in agreement with the absence of a pair-annihilation spectral feature. This was due to the rather large Thomson optical depth of the hard-Comptonization cloud, $\tau_{\rm T}\approx 5$. 

We perform spectral fits using {\sc xspec} \citep{Arnaud96}. The uncertainties given below correspond to the 90\% confidence range for a single parameter, $\Delta\chi^2 = + 2.71$. In order to assure that some local minima are not missed, we also scan the parameter space using the {\tt steppar} command of {\tt xspec}.

\subsection{The \nustar and \integral data alone}
\label{NI}

We first re-fit the \nustar and \integral data alone. We initially fit the data with the same hybrid-Comptonization model as in \zdz (where these data are also denoted as epochs 1 and 2). The model includes two Comptonization regions with two corresponding reflection regions, see above and section 3.1 of \citet{Zdziarski21b} for a detailed description. In particular, the BH is assumed to be maximally rotating, with the dimensionless spin parameter of 0.998, in which case the ISCO radius is at $\approx\! 1.24 R_{\rm g}$. That model fits very well the high-energy data of the SPI and ISGRI up to several hundred keV. In the fits, the two reflection zones have very different characteristic ionization parameters, $\lg \xi \sim 3.5$ and $\lesssim 1.7$, respectively (\citealt{Zdziarski21b}; \zdz).

As in \zdz, we allow for a residual uncertainty of the overall slope of the spectral calibration, by multiplying the model spectrum by $K E^{-\Delta\Gamma}$, where we fix $K=1$ and $\Delta\Gamma=0$ for the module A of \nustar. The fitted exponent $\Delta\Gamma$ was found in \zdz and here to be relatively small, $\lesssim\! 0.01$ for \nustar module B and for the \integral spectra. Also, we require the fitted inclination to be within the joint range of those found for the binary and the jet, $59\degr\leq i \leq 81$ \citep{Wood21, Torres20}.

In the model, the soft Comptonization is purely thermal (with a non-thermal electron tail possible, but not affecting the parameters, see \zdz), whose incident spectrum and its relativistic reflection are modelled using {\tt reflkerr} \citep{Niedzwiecki19}. The hard Comptonization is hybrid, by electrons with a predominantly Maxwellian distribution with a power-law electron tail above certain energy, modelled with {\tt reflkerr\_bb} (\zdz). In both routines, the incident Comptonization spectrum is modelled using {\tt compps} \citep{PS96}. The temperature of the seed photons is kept at 0.2\,keV, following the results of fitting \nicer data by \citet{Wang20_HXMT}. The current version of {\tt reflkerr\_bb} has a small increase in the accuracy of the non-thermal tail and an improvement of the mapping of the used reflection spectra of {\tt xillver} \citep{GK10, Garcia18} on the grid of Comptonization spectra with respect to the version used in \zdz. Still, this has only a minor effect on the fitted parameters and $\chi^2$. This model yields $\chi_\nu^2\approx 954/792$ and $1762/1307$ (which are very similar to those in \zdz) for epochs 1 and 2, respectively. 

However, since the \hxmt LE data (which we fit below) extend down to 2\,keV, we include a disk blackbody component to the model. This accounts for the observed spectral softening at low energies (cf.\ fig.\ 7 in \citealt{Zdziarski21b}). We use the {\tt diskbb} model \citep{Mitsuda84}. This significantly improves the fits with $\chi^2\approx 940/790$ and $1680/1305$ for epochs 1 and 2, respectively.

Also, we add an unblurred reflection to the above model. This is motivated by the maximum radius for which reflection is calculated in either {\tt reflkerr} or {\tt relxill} being equal to $10^3 R_{\rm g}$. At this radius, there is still a measurable broadening of the Fe K line. We first add a Gaussian Fe K$\alpha$ line. For epoch 1 and 2, it yields $\chi_\nu^2\approx 920/789$ and $1634/1304$, respectively i.e., the fit improvements are significant. We find the line is required to have the centroid energy of 6.40\,keV and the width $\ll\! 1$\,keV (indicating its origin in a cold and neutral medium), which we then fix in the fit. We have then added instead a remote, static, reflection using the {\tt hreflectnth} model \citep{Niedzwiecki19} with the same parameters of the electron distribution as the hard hybrid Comptonization (modelled by {\tt reflkerr\_bb}) but with the ionization parameter fixed at the minimum value possible in this model of $\xi=1$\,erg\,cm\,s$^{-1}$ (i.e., for the remote reflector being nearly neutral). This results in virtually no difference of the fitted parameters and $\chi_\nu^2$ with respect to the model with a Gaussian line for epoch 2. This is because only the Fe K$\alpha$ line contributes to the model significantly in {\tt hreflectnth}. Thus, in order to keep the model as simple as possible, we use a Gaussian component hereafter. Our model is then
\begin{align}
&{\tt plabs*tbabs(diskbb+reflkerr+reflkerr\_bb}\nonumber\\
&\qquad{\tt +gauss)}.\label{reflkerr}
\end{align}

Furthermore, \bei considered a variable irradiation index, $q$, corresponding to the disk irradiating flux of $\propto R^{-q}$. In the models above, the standard value (corresponding either to the disk dissipation, \citealt{SS73} or irradiation by a central point source) of $q=3$ was used. However, we find that for the present data $q$ is only weakly constrained to $\geq$2.2 for epoch 1, and allowing it to be free does not improve the $\chi^2$. For epoch 2, we obtain a similar $q\approx 3.0^{+1.5}_{-0.8}$ (and no $\chi^2$ improvement). Thus, we hereafter keep the fixed $q=3$. Our fitting results for this case are given in Table \ref{t_reflkerr}. Following \citet{Zdziarski21b} and \zdz, we have allowed the values of $i$ and $Z_{\rm Fe}$ to be different for the two epochs. The main motivation here is to see the effect of different data sets and models for the fitted values. In our view, X-ray spectral models cannot by themselves be taken as fully reflecting the physical reality. They provide only some approximations, which then can be tested by combining with results from timing and other studies (see, e.g., \citealt{DeMarco21}). We also note that while the Fe abundance has to remain constant, some changes of the fitted inclinations are possible due to precession and warping of various parts of the reflecting disk. 

The results in Table \ref{t_reflkerr} for epoch 2 are relatively similar to those obtained by \zdz without including the disk blackbody and narrow Fe K line components. We can see that the fitted truncation radius decreased, from $31^{+9}_{-5} R_{\rm g}$ to $17^{+3}_{-5} R_{\rm g}$, and the Fe abundance slightly increased. On the other hand, the present fit for epoch 1 gives $i\approx 59\degr$, lower than that in \zdz. As noted above, we constrained $i$ to $\geq\! 59\degr$. When this constraint is relaxed, the best fit $i$ still remains at $59\degr$. Then, the Fe abundance has significantly increased, to a rather unlikely (at the best fit) value of $Z_{\rm Fe}\approx 3.6^{+0.5}_{-1.7}$. This is apparently an artefact of the fitted model. As discussed by \citet{Garcia18n}, the reflector density being kept at a fixed low value in the used reflection model may lead to $Z_{\rm Fe}$ being in some cases artificially high.

We have then studied alternative models. For the sake of simplicity, we consider here only epoch 2, which has a much longer exposure than epoch 1, see Table \ref{log}. Above, the reflection of the soft components came from an inner part of the disk and was much more strongly ionized than that of the hard components, which came from an outer disk part, see fig.\ 4 in \citet{Zdziarski21b}. We have thus three alternative possibilities. The reflection of the hard component could be still from an outer part but with a high ionization, or it could be from an inner part with either low or high ionization. We have found that all three possibilities correspond to local minima with the values of $\chi^2$ significantly higher than our previous spectral solution with the hard component reflecting from an outer part with low ionization (representing thus the global minimum). The values of $\Delta \chi^2$ are $+66$, $+33$ and $+21$, respectively. Summarizing our investigations above, our chosen hybrid-Compton model to broad-band spectra is most likely among the considered models based on the $\chi^2$ criterion. 

On the other hand, the main purpose of this work is to study the \hxmt spectral data (alone and together with those from \nustar and \integral) for this source, which useful energy range is (as we find below) $\lesssim$150\,keV. Furthermore, we would like to be able to compare our above results, obtained with the {\tt reflkerr} family of models, with those of the {\tt relxillCp} models \citep{GK10,Dauser16,Garcia18}. Since {\tt relxillCp} does not have an option to include a non-thermal tail in the electron distribution, we limit the energy range of the fitted data to $\leq$150\,keV, which allows us not to include the high-energy electron tail, which was found to be necessary to fit the \integral data at higher energies. Also, given the structure of {\tt relxillCp}, we have to consider the variant of that model with two Comptonization zones and the two corresponding reflection regions overlapping rather than adjacent, i.e., with two independent inner reflection radii, in spite of the objections to their physical reality discussed above. As shown in \citet{Zdziarski21b}, see their table 2, this still has a relatively little effect on the fitted parameters. Our model is then 
\begin{align}
&{\tt plabs*tbabs(diskbb+relxillCp_{\rm s}+relxillCp_{\rm h}}\nonumber\\
&\qquad{\tt +gauss)},\label{relxill}
\end{align}
where the indices `s' and `h' correspond to the soft and hard thermal Comptonization components, respectively. Note that the low-energy spectral index, $\Gamma$, is constrained to $\geq$1.20 in {\tt relxillCp}.

\begin{figure}
\centerline{
\includegraphics[width=7.7cm]{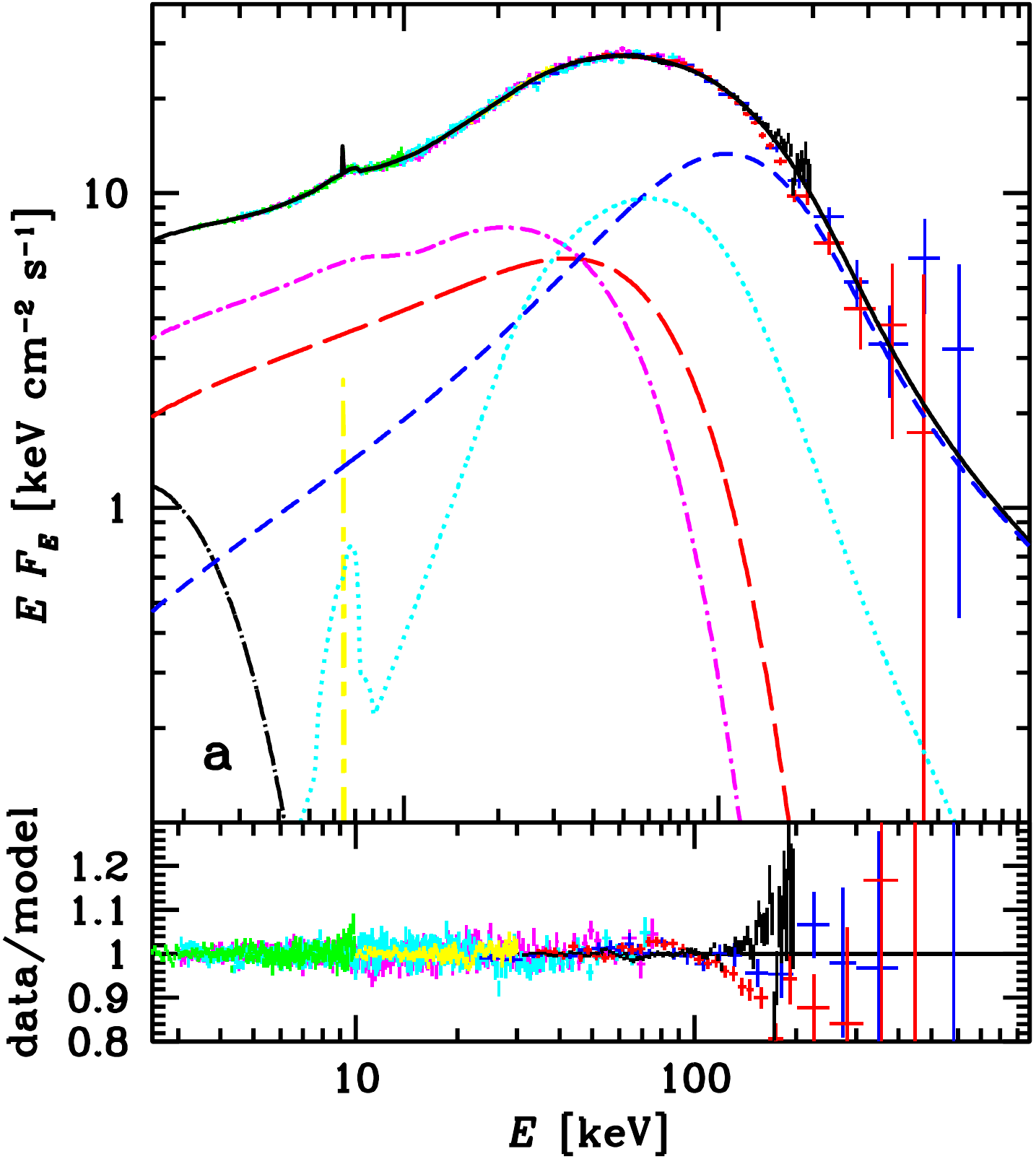}}
\centerline{
\includegraphics[width=7.7cm]{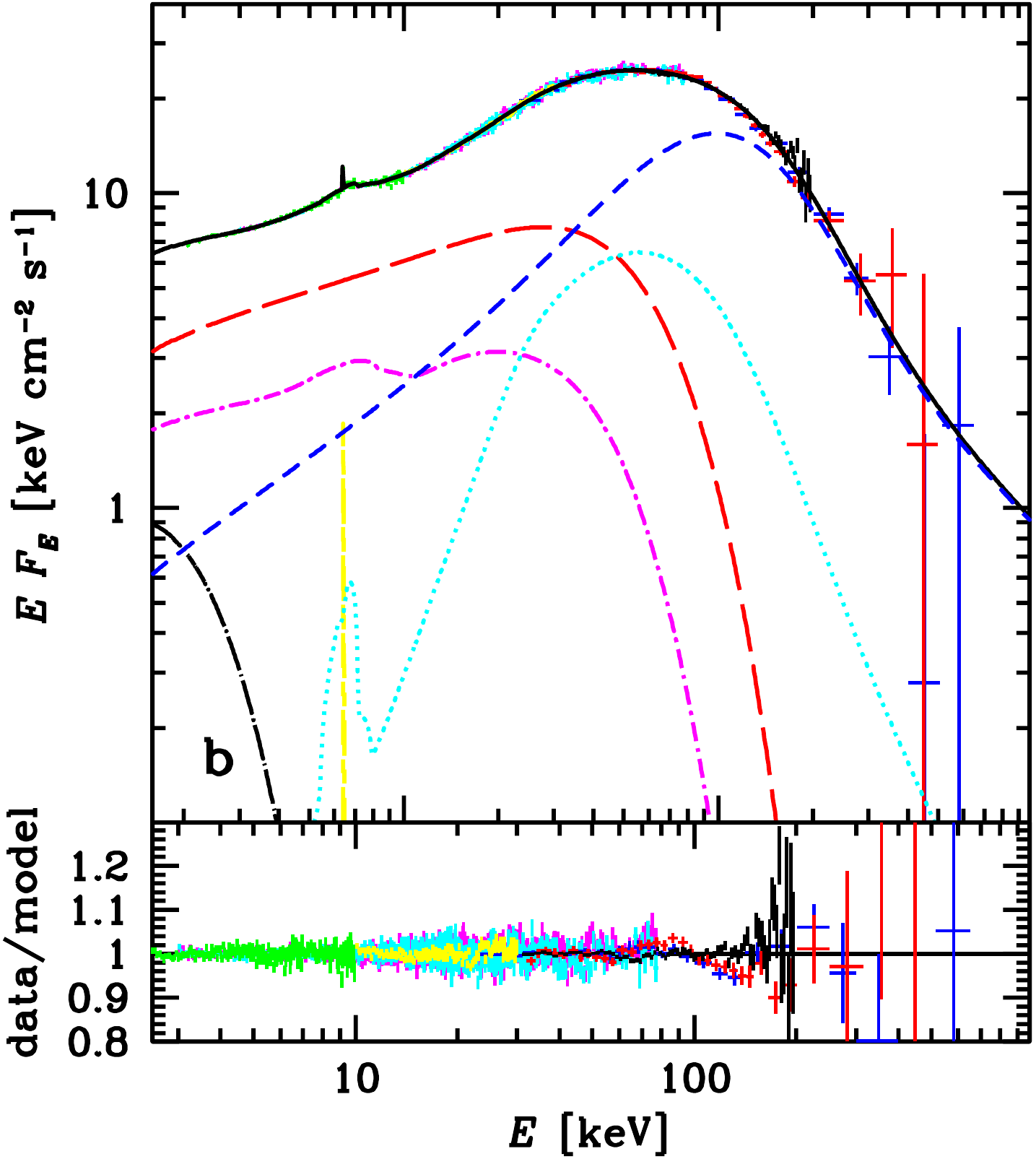}}
  \caption{The unfolded spectra and data-to-model ratios of the epochs (a) 1 and (b) 2, fitted with the model of Equation (\ref{reflkerr}), see Table \ref{t_reflkerr_full} for the parameters. The error bars show the data from \nustar A and B (magenta and cyan, respectively), SPI (blue), ISGRI (red), \hxmt LE (green), ME (yellow) and HE (black). The spectra are normalized to the \nustar A one. The softer (red long dashes) and harder (blue short dashes) Comptonization components are for thermal and hybrid electrons, respectively. The corresponding reflection components are shown by the magenta dot-dashes and cyan dots, respectively. The disk blackbody and the narrow Fe K$\alpha$ lines are shown by the black dot-dashed and yellow dashed lines, respectively. No rebinning has been applied to the plots. 
}\label{eeuf_ratio_reflkerr}
\end{figure}

The results of the fitting with this model are given in Table \ref{t_relxill}. We see that we still obtain fits where the disk is significantly truncated, and the truncation radii for the two reflectors are larger for both epochs (except for $R_{\rm in,s}$ for the reflection of the soft Comptonization component for epoch 1, which is almost the same as the corresponding one for the {\tt reflkerr} model) than the inner truncation radii obtained with the {\tt reflkerr} model (of Equation \ref{reflkerr}), see Table \ref{t_reflkerr}. The differences in the parameters with respect to the {\tt reflkerr} models are due to the differences in the models; in particular, {\tt relxillCp} uses the purely thermal Comptonization model of \citet{ZJM96} rather than the hybrid one of \citet{PS96}. Still, the two sets of results are relatively similar. 

We note that the increase of $R_{\rm in}$ from epochs 1 to 2 in Table \ref{t_relxill} is associated with a decrease of the spectral index of the dominant hard component, $\Gamma_{\rm h}$. This is expected if the cooling of this component is reduced due to fewer disk blackbody photons impinging on that component. On the other hand, we see that the electron temperature decreases, contrary to the above expectation. This shows that the energy balance is more complex, most likely involving changing the optical depth and cooling by synchrotron photons. We note that a similar lack of a correlation between $T_{\rm e}$ and the spectral hardness is seen in Cyg X-1, see figures 4 and 6 in \citet{Ibragimov05}. On the other hand, an analogous increase of $R_{\rm in}$ from epochs 1 to 2 is much weaker in the model using {\tt reflkerr}, see Table \ref{t_reflkerr}. This demonstrates a significant model dependence of the detailed fitting results.

\subsection{The \nustar, \integral and \hxmt data}
\label{n_i_hxmt}

We then include the \hxmt data in our spectral models. We fit the LE, ME and HE data in the 2--10, 10--30 and either the full range of 30--200\,keV when using Equation (\ref{reflkerr}) and 30--150\,keV with Equation (\ref{relxill}). We allow $K$ and $\Delta\Gamma$ to be free for the LE and ME data; for HE, we found that $\Delta\Gamma$ is compatible with null for epoch 1, and allowing it free does not improve the fit. We thus keep it fixed at null, and allow only $K$ to be free. 

Our results are given in Tables \ref{t_reflkerr_full} and \ref{t_relxill_full} and Figure \ref{eeuf_ratio_reflkerr}. We find that the addition of the \hxmt data only moderately changes our results. In particular, the reflecting disk is substantially truncated for all of the fits during both epochs. In all cases, we find $R_{\rm in}\gtrsim 10 R_{\rm g}$. However, we see that using the present model and including the \hxmt data results in the reflection components being quite strong for epoch 1, see Figure \ref{eeuf_ratio_reflkerr}(a), while no such issue appeared for the \nustar+\integral data alone fitted in \zdz. 

We have then tested again the effect of allowing the irradiation index, $q$, to be free. While it gave no fit improvement at all in the case of the \nustar+\integral data alone, we find it now only slightly, by a few, decrease the values of $\chi^2$, while the fitted values are $q\gtrsim 2$. Still, those models have similarly large values of the disk truncation radii, and show no significant changes in the parameters. We thus have not investigated those models further. 

We clearly see apparently spurious high-energy tails in the HE data above $\sim$100\,keV. The presence of those upturns have resulted in hardenings of the best-fit models, and in consequence the ISGRI and SPI data are below the model at $\gtrsim$100\,keV. Since the calibration of the \integral detectors is rather well established, with mutual agreement of the spectra from its three high-energy detectors (including the PICSiT), as shown for \source in \zdz, the actual departure of the high-energy part of the HE spectrum corresponds to ratio of the HE and \integral spectra, and it is higher than the departures from the present best fits. 

While we find also low value of $\Delta\Gamma$ for the fitted LE and HE spectra, we find the rather high values of the best fit $\Delta\Gamma\approx 0.05$, 0.06 for the ME data for epochs 1 and 2, respectively\footnote{This problem appears to be specific to the \hxmt data for \source; no such a large $\Delta\Gamma$ is found when fitting the \hxmt spectra of either Crab or the BH X-ray binary MAXI J1535--571.}.

\begin{table}
\caption{The results of spectral fitting the \nustar+\integral+\hxmt data in the 2--650\,keV energy range
}
   \centering\begin{tabular}{ccccc}
\hline
Component & Parameter & Epoch 1 & Epoch 2 \\
\hline
ISM absorption & $N_{\rm H}$ $[10^{21}]$\,cm$^{-2}$ & \multicolumn{2}{c}{1.4f}\\
\hline
Joint constraints & $i$ $[\degr$] & $59^{+3}_{-0}$ & $60^{+1}_{-0}$\\
& $Z_{\rm Fe}\, [\sun]$ & $2.3^{+0.3}_{-0.1}$ & $2.1^{+0.1}_{-0.2}$\\
\hline
Disk & $kT_{\rm in}$ & $0.50^{+0.02}_{-0.01}$ & $0.50^{+0.02}_{-0.02}$\\
and a narrow & $N_{\tt diskbb}\,[10^{3}]$ & $3.3^{+0.6}_{-0.6}$ & $2.6^{+0.7}_{-0.4}$\\
6.40\,keV line & $N_{{\rm Fe\,K}\alpha}\,[10^{-3}]$ & $3.1^{+0.1}_{-0.1}$ & $2.3^{+0.4}_{-0.3}$\\
\hline
Thermal  & $y_{\rm th}$ & $0.71^{+0.01}_{-0.01}$ & $0.61^{+0.01}_{-0.01}$\\
Comptonization &$\Gamma_{\rm th}$ & $1.62^{+0.01}_{-0.01}$ & $1.71^{+0.01}_{-0.01}$ \\
and reflection & $kT_{\rm e,th}$ [keV] & $15.5^{+0.4}_{-0.5}$ & $12.4^{+0.3}_{-0.2}$\\
&$R_{\rm in}\, [R_{\rm g}]$ & $15^{+1}_{-4}$ & $15^{+1}_{-3}$\\
& ${\cal R}_{\rm th}$ & $1.40^{+0.05}_{-0.04}$ & $0.49^{+0.02}_{-0.05}$\\
& $\log_{10} \xi_{\rm th}$ & $4.20^{+0.03}_{-0.01}$ & $3.88^{+0.02}_{-0.04}$\\
& $N_{\rm th}$ & $1.62^{+0.50}_{-0.06}$ & $2.67^{+0.18}_{-0.07}$\\
\hline 
Hybrid &$y_{\rm h}$ & $1.38^{+0.02}_{-0.05}$ & $1.03^{+0.03}_{-0.03}$\\
Comptonization &$\Gamma_{\rm h}$  & $1.19^{+0.02}_{-0.01}$ & $1.27^{+0.02}_{-0.02}$\\
and reflection& $kT_{\rm e,h}$ [keV] & $27.5^{+0.3}_{-2.4}$ & $23.1^{+2.2}_{-0.4}$\\
&$\gamma_{\rm min}$ & $1.26^{+0.16}_{-0.02}$ & $1.15^{+0.25}_{-0.02}$\\
&$p$ & $3.66^{+0.10}_{-0.08}$ & $3.7^{+0.3}_{-0.1}$\\
&$\Delta R\, [R_{\rm g}]$ & $17^{+4}_{-6}$ & $27^{+5}_{-6}$\\ 
& ${\cal R}_{\rm h}$ & $1.29^{+0.01}_{-0.02}$ & $0.73^{+0.09}_{-0.02}$\\
& $\log_{10} \xi_{\rm h}$ & $0^{+1.7}$ & $0^{+1.2}$\\
& $N_{\rm h}$ & $0.35^{+0.03}_{-0.01}$ & $0.46^{+0.01}_{-0.01}$\\
\hline
LE/\nustar\,A & $\Delta\Gamma\,[10^{-2}]$ & $0.9^{+0.1}_{-0.6}$ & $1.0^{+0.5}_{-0.5}$\\
ME/\nustar\,A & $\Delta\Gamma\,[10^{-2}]$ & $5.0^{+0.2}_{-0.5}$ & $6.2^{+0.4}_{-0.4}$\\
\hline
& $\chi_\nu^2$  & 1684/1192 & 2210/1706\\
\hline
\end{tabular}
\tablecomments{The {\tt reflkerr} model, Equation (\ref{reflkerr}), is used. }
\label{t_reflkerr_full}
\end{table}

\begin{table}
\caption{The results of spectral fitting the \nustar+\integral+\hxmt data in the 2--150\,keV energy range
}
   \centering\begin{tabular}{ccccc}
\hline
Component & Parameter & Epoch 1 & Epoch 2 \\
\hline
ISM absorption & $N_{\rm H}$ $[10^{21}]$\,cm$^{-2}$ & \multicolumn{2}{c}{1.4f}\\
\hline
Joint constraints & $i$ $[\degr$] & $59^{+6}_{-0}$ & $62^{+3}_{-3}$\\
& $Z_{\rm Fe}\, [\sun]$ & $1.9^{+0.2}_{-0.1}$ & $1.8^{+0.2}_{-0.2}$\\
\hline
Disk & $kT_{\rm in}$ & $0.48^{+0.03}_{-0.03}$ & $0.48^{+0.02}_{-0.03}$\\
and a narrow & $N_{\tt diskbb}\,[10^{3}]$ & $4^{+2}_{-2}$ & $3^{+1}_{-1}$ \\
6.40\,keV line & $N_{{\rm Fe\,K}\alpha}\,[10^{-3}]$ & $3^{+1}_{-1}$ & $2.0^{+0.4}_{-0.4}$ \\
\hline
Soft  &$\Gamma_{\rm s}$ & $1.71^{+0.01}_{-0.01}$ & $1.71^{+0.01}_{-0.01}$ \\
Comptonization & $kT_{\rm e,s}$ [keV] & $14^{+1}_{-2}$ & $13^{+1}_{-1}$\\
and reflection &$R_{\rm in,s}\, [R_{\rm g}]$ & $13^{+4}_{-3}$ & $70^{+47}_{-30}$\\
& ${\cal R}_{\rm s}$ & $0.69^{+1.31}_{-0.04}$ & $0.7^{+0.5}_{-0.3}$\\
& $\log_{10} \xi_{\rm s}$ & $3.95^{+0.30}_{-0.04}$ & $3.9^{+0.1}_{-0.1}$\\
& $N_{\rm s}\,[10^{-2}]$ & $6.4^{+0.1}_{-0.1}$ & $6.3^{+0.5}_{-0.4}$\\
\hline 
Hard  &$\Gamma_{\rm h}$  & $1.31^{+0.03}_{-0.04}$ & $1.21^{+0.03}_{-0.01}$\\
Comptonization& $kT_{\rm e,h}$ [keV] & $33^{+2}_{-2}$ & $29^{+1}_{-1}$\\
and reflection
&$R_{\rm in,h}\, [R_{\rm g}]$ & $41^{+11}_{-9}$ & $54^{+15}_{-10}$\\ 
& ${\cal R}_{\rm h}$ & $1.0^{+0.4}_{-0.2}$ & $1.0^{+0.2}_{-0.5}$\\
& $\log_{10} \xi_{\rm h}$ & $0^{+2.3}$ & $0^{+2.0}$\\
& $N_{\rm h}\,[10^{-2}]$ & $6.6^{+0.6}_{-0.4}$ & $5.7^{+0.3}_{-0.1}$\\
\hline
LE/\nustar\,A & $\Delta\Gamma\,[10^{-2}]$ & $0.9^{+0.6}_{-0.6}$ & $1.0^{+0.5}_{-0.5}$\\
ME/\nustar\,A & $\Delta\Gamma\,[10^{-2}]$ & $5.0^{+0.6}_{-0.6}$ & $6.1^{+0.5}_{-0.5}$\\
\hline
& $\chi_\nu^2$  & 1561/1158 & 2139/1675\\
\hline
\end{tabular}
\tablecomments{The {\tt relxillCp} model, Equation (\ref{relxill}), is used. }
\label{t_relxill_full}
\end{table}

Then, following the recommendation of the \hxmt team, we have tested the effect of ignoring the ME data in the 21--24\,keV range. However, we have found that including those data has virtually no effect on the fits, and results only in a modest increase of $\chi^2_\nu$. 

\begin{table}[t!]
\caption{The results of spectral fitting the \hxmt data only in the 2--150\,keV energy range
}
   \centering\begin{tabular}{ccccc}
\hline
Component & Parameter & Epoch 1 & Epoch 2 \\
\hline
ISM absorption & $N_{\rm H}$ $[10^{21}]$\,cm$^{-2}$ & \multicolumn{2}{c}{1.4f}\\
\hline
Joint constraints & $i$ $[\degr$] & $63^{+4}_{-3}$ & $59^{+8}_{-0}$\\
& $Z_{\rm Fe}\, [\sun]$ & $1.6^{+0.3}_{-0.3}$ & $2.0^{+0.4}_{-0.4}$\\
\hline
Disk & $kT_{\rm in}$ & $0.46^{+0.02}_{-0.02}$ & $0.45^{+0.05}_{-0.05}$\\
and a narrow & $N_{\tt diskbb}\,[10^{3}]$ & $4.8^{+2.3}_{-0.8}$ & $4.4^{+3.2}_{-1.7}$ \\
6.40\,keV line & $N_{{\rm Fe\,K}\alpha}\,[10^{-3}]$ & $2.6^{+0.7}_{-0.9}$ & $1.7^{+0.8}_{-0.8}$ \\
\hline
Soft  &$\Gamma_{\rm s}$ & $1.73^{+0.03}_{-0.03}$ & $1.75^{+0.02}_{-0.02}$ \\
Comptonization & $kT_{\rm e,s}$ [keV] & $13^{+1}_{-1}$ & $13^{+1}_{-1}$\\
and reflection &$R_{\rm in,s}\, [R_{\rm g}]$ & $230^{+770}_{-175}$ & $1.4^{+7.4}_{-0.2}$\\
& ${\cal R}_{\rm s}$ & $0.5^{+0.1}_{-0.2}$ & $0.5^{+0.6}_{-0.1}$\\
& $\log_{10} \xi_{\rm s}$ & $4.0^{+0.1}_{-0.1}$ & $3.6^{+0.3}_{-0.1}$\\
& $N_{\rm s}\,[10^{-2}]$ & $7.1^{+0.6}_{-0.3}$ & $6.9^{+0.5}_{-1.3}$\\
\hline 
Hard  &$\Gamma_{\rm h}$  & $1.27^{+0.05}_{-0.07}$ & $1.27^{+0.07}_{-0.06}$\\
Comptonization& $kT_{\rm e,h}$ [keV] & $32^{+1}_{-1}$ & $32^{+2}_{-2}$\\
and reflection
&$R_{\rm in,h}\, [R_{\rm g}]$ & $33^{+22}_{-8}$ & $59^{+44}_{-19}$\\ 
& ${\cal R}_{\rm h}$ & $0.5^{+0.1}_{-0.1}$ & $0.9^{+0.1}_{-0.1}$\\
& $\log_{10} \xi_{\rm h}$ & $0^{+1.8}$ & $0^{+2.7}$\\
& $N_{\rm h}\,[10^{-2}]$ & $6.1^{+0.4}_{-0.7}$ & $6.1^{+0.06}_{-0.04}$\\
\hline
LE/HE & $\Delta\Gamma\,[10^{-2}]$ & $-1.4^{+1.2}_{-1.4}$ & $-0.7^{+1.3}_{-0.3}$\\
ME/HE & $\Delta\Gamma\,[10^{-2}]$ & $4.8^{+0.2}_{-1.0}$ & $6^{+0}_{-1}$\\
\hline
& $\chi_\nu^2$  & 552/370 & 451/370\\
\hline
\end{tabular}
\tablecomments{The {\tt relxillCp} model, Equation (\ref{relxill}), is used. The values of $R_{\rm in}$ obtained here appear spurious, as they are completely different from those obtained using the \nustar data, see Tables \ref{t_reflkerr}--\ref{t_relxill_full}}.
\label{t_relxill_hxmt}
\end{table}

\subsection{\hxmt data alone}
\label{hxmt}

We have then fit the \hxmt data alone. For the sake of simplicity, we fit here only the {\tt relxillCp} model, Equation (\ref{relxill}). Given the results of the joint fitting, we assume $\Delta\Gamma=0$, $K=1$ for the HE data. However, we have found that allowing free values of $\Delta\Gamma$ for the LE and ME leads to rather large, and obviously spurious, results. We thus have constrained their absolute values to be within the uncertainty ranges of the corresponding joint fits, see Table \ref{t_relxill_full}. Our results are given in Table \ref{t_relxill_hxmt}. 

We can see that these data alone poorly constrain the inner radius. It is found very large for epoch 1, and very small for epoch 2. This is opposite to the behavior found for the fits with the \nustar+\integral data alone, Table \ref{t_relxill}, and those including the \hxmt data, Table \ref{t_relxill_full}. In those cases, the inner radius for epoch 2 was found to be several times that for epoch 1, with both showing significant disk truncation. This particular \hxmt result is clearly spurious. It appears that the \hxmt data are, unfortunately, not suitable to determine the truncation radii during the evolution of \source.

\section{Discussion}
\label{discussion}

We have found a good agreement between the overall shapes of the \hxmt data and those of \nustar and \integral. The two main outstanding issues are the slope of the ME data requiring a relatively large correction to bring it into an agreement with the \nustar data, and the presence of apparently spurious high-energy tails in the HE data at $E\gtrsim 10^2$\,keV. We note, however, that \citet{Li20} estimate the systematic error of the HE data at 150\,keV to be as high as 5\%, whose inclusion would strongly reduce the statistical significance of the tail.

Our major robust result is yet another confirmation of relatively large inner truncation radii of the reflecting disk in \source, $R_{\rm in}\gtrsim 10 R_{\rm g}$, for both the \nustar data alone and those including \hxmt. On the other hand, the fits of a large set of \hxmt data in \bei\ {\it assumed\/} that the disk extends down to the ISCO of a maximally rotating BH, i.e., $R_{\rm in}\approx 1.24 R_{\rm g}$. This disk was irradiated by a corona above it (see equation 1 in \bei) with a free irradiation index, with the best-fit values of $q\sim 0$. Such low values of $q$ are possible only if either the irradiation of innermost parts of the disk is strongly reduced with respect to the standard coronal dissipation profile \citep{NT73} or those disk parts are strongly fragmented. Alternatively, such values of $q$ may mimic the effect of the zero-stress boundary condition at the ISCO for a slowly-spinning BH, where the peak of the dissipation profile is at $>10 R_{\rm g}$. In fact, the parameters $q$ and $R_{\rm in}$ are strongly correlated with each other \citep{Wilkins12}, and often cannot be independently determined.

In the case of the highly accurate \nustar data, we have found no fit improvement at all when allowing a free $q$, and the full compatibility of the models with the standard value of $q=3$. When the \hxmt data are added, we find slight fit improvements only, and $q\gtrsim 2$. This difference with respect to the results of \bei is due to (1) the inner disk assumed at the ISCO in \bei while estimated from the spectral fits in the present work, and (2) taking into account the spectral complexity of this source in the present work, with the presence of two Comptonization regions, the softer one dominating at softer X-rays, and a harder one at harder X-rays. Such stratification is also indicated by the presence of hard lags in this source, i.e., harder X-rays delayed with respect to softer ones \citep{DeMarco21}, and by spectral-timing modelling of \citet{Kawamura21}. Spectral complexity of this source was also found by \citet{Buisson19}, who fitted the \nustar spectra with reflection of two lampposts at different heights. The necessity of spectral complexity in \source is discussed in detail by \citet{Zdziarski21b}. 

Furthermore, the widths of the Fe K complexes fitted for the \hxmt data alone were found significantly different from the fits to the combined \nustar and \hxmt data. As found in Section \ref{hxmt}, the \hxmt data alone did show the reflecting disk at the ISCO for a maximally rotating BH for epoch 2, opposite to the case of either the \nustar or combined data sets, Sections \ref{NI} and \ref{n_i_hxmt}. On the other hand, the truncation radius for epoch 1 was found to be very large (Table \ref{t_relxill_hxmt}). This comparison indicates that the \hxmt data for \source alone are not suitable for detailed X-ray spectroscopy of the Fe K region. On the other hand, the \hxmt data together with those of \nustar do show moderate truncation radii for both epochs and using two different models.

The inner disk temperatures we have found are $kT_{\rm in}\approx 0.4$--0.5\,keV in all cases, which is significantly higher than the values obtained by fitting the NICER data, which are $\approx$0.2\,keV \citep{Wang20_HXMT}. This indicates the presence of a further softening of the apparent Comptonization continuum toward lower energies, see figure 7 in \citet{Zdziarski21b} rather the presence of two separate disk blackbody components. Then, our usage of the disk blackbody model should be considered to be a phenomenological description only of inhomogeneous Comptonization/reflection spectra. This is indicated by the apparently complex structure of the accretion flow in the hard state, implied, e.g., by the results of the frequency-resolved spectral fitting of \nicer data by \citet{Axelsson21} and \citet{Dzielak21}, see also \citet{Kawamura21}. 

We can also compare our spectral fitting results to some analytical estimates of truncation radius. First, we can use the normalization of the disk blackbody spectral component. This normalization was found at $N_{\tt diskbb}\approx 3$--$4\times  10^3$ in Tables \ref{t_reflkerr_full}--\ref{t_relxill_hxmt}. The implied $R_{\rm in}$ depends on the color correction, $\kappa$, see, e.g., equation (8) in \citet{Zdziarski21}. This correction has been found to be within $\kappa\approx 1.3$--1.7 \citep{Davis05}. Then, for $D=3$\,kpc, $M=7\msun$ and $i=65\degr$, we find $R_{\rm in}/R_{\rm g}$ within $\approx$6--13, which is somewhat below our fitted values for joint data. However, we can also estimate the blackbody temperature resulting from irradiation of the disk close to $R_{\rm in}$ by the Comptonization emission. The irradiating Comptonization fluxes of the inner, soft, component for epochs 1 and 2 are $F_{\rm C}\approx 3$--$4\times 10^{-8}$\,erg\,cm$^{-2}$\,s$^{-1}$. Absorption of a part $(1-a)$ of this flux, where $a$ is the back-scattering albedo, results in a quasi-blackbody emission at a temperature above the effective one, see \citet{ZDM20}. This irradiating temperature is estimated in equation (9) of \citet{Zdziarski21}. Typical values of $a$ for ionized reflecting media were estimated to be within $a\approx 0.3$--0.7 \citep{ZDM20}. We also neglect here the reduction of the observed Comptonization flux with respect to that irradiating the disk due to scattering in the hot medium, which is a conservative assumption. Then that formula gives the color temperature due to irradiation alone as $kT_{\rm in,irr}\approx 0.6$--0.9\,keV, i.e., more than the fitted inner temperatures. This shows that the assumption that the fitted {\tt diskbb} corresponds to actual disk blackbody is unphysical, in agreement with the finding of a lower inner temperature in the NICER data \citep{Wang20_HXMT}. Finally, we estimate the inner radius at which irradiation would result in a given inner temperature, equation (6) in \citet{Zdziarski21}. This gives relatively large values of $R_{\rm in}\gtrsim 50 R_{\rm g}$. Thus, our estimates favor strong truncation.

As we noted above, \bei interpreted their fits of \hxmt data as showing the presence of an outflowing corona. A strong argument for it was a decrease of the fitted reflection fraction with the decreasing spectral hardness, which was, in turn, correlated with the shortening time lags of soft X-rays with respect to hard ones (measured by \nicer). This discovery was interpreted as due to the Doppler de-boosting of the disk irradiation by a corona whose velocity decreases with the height. Note that the values of ${\cal R}\ll 1$ found by them were due to a bug in previous versions of {\tt relxill}\footnote{At the time of preparing the publication of \bei, {\tt relxill} v.\ 1.4.0 was the latest available version, with the bug fixed in v.\ 1.4.3, see \url{http://www.sternwarte.uni-erlangen.de/~dauser/research/relxill/}. This version has been used in the present work.}. After that corrections, the fitted values of ${\cal R}$ increase by a factor of $\approx$3, but the trend of the decreasing reflection fraction remains present (You \& Zdziarski, in preparation).

Still, it is possible that the hard Comptonization component originates in an outflowing corona forming an inner part of the jet (as noticed by \citealt{Zdziarski21b}). This was also suggested by \citet{Ma21} based on the behavior of X-ray quasi-periodic oscillations. Then, bulk-motion Comptonization may provide another spectral component (e.g., \citealt{Mondal21}). However, including this physical process is beyond the scope of our paper.

Finally, we comment on the relatively low values of the fitted electron temperatures for fits with hybrid Comptonization, Tables \ref{t_reflkerr} and \ref{t_reflkerr_full}. As discussed in \zdz, this may be due either to the onset of pair production by photons scattered by the non-thermal tail \citep{Coppi99, Fabian17} or due to energy balance in the flow emitting synchrotron photons \citet{PV09,MB09}. Still, \source in the hard state fitted by hybrid Comptonization shows lower values of $kT_{\rm e}$ than Cyg X-1 fitted by the same spectral model \citep{McConnell02}, which may be due to the disk truncation being stronger in Cyg X-1 than in \source, which effect may in turn be related to a relatively low Eddington ratio in Cyg X-1. Indeed, there is ample evidence for the truncation radius in the hard state to decrease with the increasing luminosity, see, e.g., \citet{Plant15} and \citet{Basak16} for such behavior in the BH transient binary GX 339--4. We note, however, that since both of our chosen data sets correspond to almost identical bolometric luminosity (and the spectral shape), we cannot here study the evolution of the truncation radius with either the luminosity of the accretion rate.

\section{Conclusions}

Our main results are as follows.

We have performed a thorough analysis of the simultaneous \nustar and \integral observations of \source in its hard spectral state. We have extended the original analysis in \zdz, and found that addition of a disk blackbody and unblurred reflection significantly improves the fits. We have tested source geometries alternative to those used in \zdz, but found they provide worse fits. We have also fitted models using the {\tt relxill} X-ray spectroscopy model, and found it yields similar results to our original models (using {\tt reflkerr}). 

Then, we have included the simultaneous \hxmt data, and repeated our fits with {\tt reflkerr} and {\tt relxill}. We have found a good mutual agreement between the \nustar, \integral and \hxmt data, with the fits yielding similar parameters to the previous ones.

In all of the above cases, we have found the reflecting disk to be truncated, with $R_{\rm in}\gtrsim 10 R_{\rm g}$. We have also tested possible departure of the irradiation index, $q$, from the canonical value of 3. However, we have found allowing it free does not improve the fits, and the best fit values remain close to 3. This provides a confirmation for the assumed coronal geometry, with the dissipation rate in the corona following that of the disk. On the other hand, constraints from re-emission of the absorbed part of the irradiation flux yield $R_{\rm in}\gg 10 R_{\rm g}$. 

We have then fitted the \hxmt data alone. We have found that they cannot constrain the inner disk radius. Our recommendation is that constraints on $R_{\rm in}$ from other instruments are used.

\section*{Acknowledgments}
We thank the referee for valuable comments. We acknowledge support from the Polish National Science Centre under the grants 2015/18/A/ST9/00746 and 2019/35/B/ST9/03944, and from the Natural Science Foundation of China (U1931203 and 11903024). Our work also benefitted from discussions during Team Meetings in the International Space Science Institute (Bern).

\bibliography{../../allbib}{}
\bibliographystyle{aasjournal}

\label{lastpage}
\end{document}